\documentclass{article}

\usepackage{epsfig}
\begin{document}

\title{\Large \bf QED Radiative Corrections to Asymmetries of Elastic
ep-scattering in Hadronic Variables}

\author{A.V. Afanasev$^{a)}$, I. Akushevich$^{a)}$\footnote{on leave
of absence from the National
Center of Particle and High Energy Physics,
220040 Minsk, Belarus}, A.Ilyichev$^{b)}$, N.P.Merenkov$^{c)}$ }
\maketitle
\begin{center}
{\small {\it $^{(a)}$ North Carolina Central University,
Durham, NC 27707, USA \\ and  \\
Jefferson Lab, Newport News, VA 23606, USA\\}}
{\small {\it $^{(b)}$ National
Center of Particle and High Energy Physics,
220040 Minsk, Belarus \\}}
{\small {\it{$^{(c)}$ NSC "Kharkov Institute of Physics and Technology" \\
}}}
{\small {\it {63108, Akademicheskaya 1, Kharkov, Ukraine}}}
\end{center}

\begin{abstract}
Compact analytical formulae for QED radiative corrections in the processes
of
elastic $e-p$ scattering are obtained in the case when kinematic
variables are
reconstructed from the recoil proton momentum measured. Numerical analysis
is
presented under kinematic conditions of current experiments at JLab.
\end{abstract}

\section{Introduction}

With the advent of new-generation electron accelerators such as CEBAF,
experiments on elastic electron--proton scattering in the GeV-range
were brought to the new level of precision. One of the novel features of
the
experiments
is that in addition to the scattered electron, the momentum of recoil
proton can be directly measured, while both the electron beam and the proton
target may be polarized.
The cross section
depends only on one kinematic variable, so the measurement of final
momenta gives
several possibilities to
reconstruct this kinematic variable from measurements. It results in
different methods of data analysis, the best of them being chosen
from experimental conditions such as resolution and others. Often it is
used for
reducing influence of radiative effects, which  accompany any process with
electrons. 
Ref.\cite{Bardin} presents a review of the radiative corrections
calculations under different approaches of kinematic variable reconstructions
in experiments at HERA.

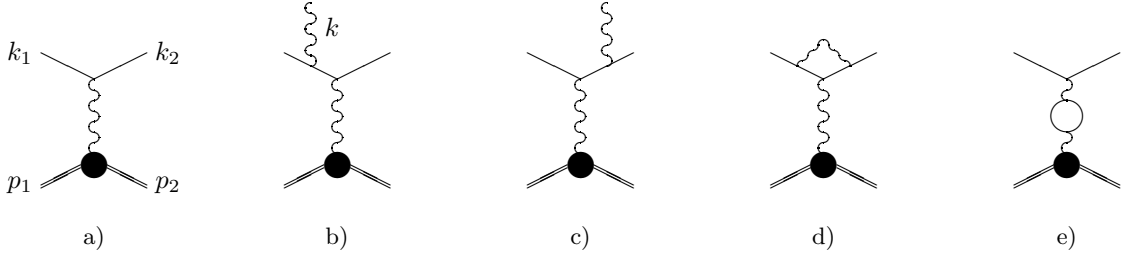
\begin{figure}[t]
\begin{tabular}{ccccc}
\begin{picture}(80,100)
\put(30,60){\line(2,-1){20.}}
\put(50,50){\line(2,1){20.}}
\put(50,17.5){\circle*{10.}}
\multiput(50,28)(0,8){3}{\oval(4.0,4.0)[r]}
\multiput(50,24)(0,8){4}{\oval(4.0,4.0)[l]}
\put(22,10){\makebox(0,0){$p_1$}}
\put(78,10){\makebox(0,0){$p_2$}}
\put(22,60){\makebox(0,0){$k_1$}}
\put(78,60){\makebox(0,0){$k_2$}}
\put(30,10){\line(2,1){15.}}
\put(30,9){\line(2,1){15.}}
\put(55,17.5){\line(2,-1){15.}}
\put(55,16.5){\line(2,-1){15.}}
\put(50,-10){\makebox(0,0){\small a)}}
\end{picture}
&
\begin{picture}(80,100)
\multiput(40,57)(0,8){3}{\oval(4.0,4.0)[r]}
\multiput(40,61)(0,8){3}{\oval(4.0,4.0)[l]}
\put(30,60){\line(2,-1){20.}}
\put(50,50){\line(2,1){20.}}
\put(48,70){\makebox(0,0){$k$}}
\put(50,17.5){\circle*{10.}}
\multiput(50,28)(0,8){3}{\oval(4.0,4.0)[r]}
\multiput(50,24)(0,8){4}{\oval(4.0,4.0)[l]}
\put(30,10){\line(2,1){15.}}
\put(30,9){\line(2,1){15.}}
\put(55,17.5){\line(2,-1){15.}}
\put(55,16.5){\line(2,-1){15.}}
\put(50,-10){\makebox(0,0){\small b)}}
\end{picture}
&
\begin{picture}(80,100)
\multiput(60,57)(0,8){3}{\oval(4.0,4.0)[r]}
\multiput(60,61)(0,8){3}{\oval(4.0,4.0)[l]}
\put(30,60){\line(2,-1){20.}}
\put(50,50){\line(2,1){20.}}
\put(50,17.5){\circle*{10.}}
\multiput(50,28)(0,8){3}{\oval(4.0,4.0)[r]}
\multiput(50,24)(0,8){4}{\oval(4.0,4.0)[l]}
\put(30,10){\line(2,1){15.}}
\put(30,9){\line(2,1){15.}}
\put(55,17.5){\line(2,-1){15.}}
\put(55,16.5){\line(2,-1){15.}}
\put(50,-10){\makebox(0,0){\small c)}}
\end{picture}
&
\begin{picture}(80,100)
\put(30,60){\line(2,-1){20.}}
\put(50,50){\line(2,1){20.}}
\put(50,17.5){\circle*{10.}}
\multiput(50,28)(0,8){3}{\oval(4.0,4.0)[r]}
\multiput(50,24)(0,8){4}{\oval(4.0,4.0)[l]}
\multiput(42,55)(4,4){3}{\oval(4.0,4.0)[lt]}
\multiput(42,59)(4,4){2}{\oval(4.0,4.0)[br]}
\multiput(50,63)(4,-4){3}{\oval(4.0,4.0)[tr]}
\multiput(54,63)(4,-4){2}{\oval(4.0,4.0)[bl]}
\put(30,10){\line(2,1){15.}}
\put(30,9){\line(2,1){15.}}
\put(55,17.5){\line(2,-1){15.}}
\put(55,16.5){\line(2,-1){15.}}
\put(50,-10){\makebox(0,0){\small d)}}
\end{picture}
&
\begin{picture}(80,100)
\put(30,60){\line(2,-1){20.}}
\put(50,50){\line(2,1){20.}}
\put(50,17.5){\circle*{10.}}
\multiput(50,48)(0,8){1}{\oval(4.0,4.0)[r]}
\multiput(50,44)(0,8){1}{\oval(4.0,4.0)[l]}
\multiput(50,28)(0,8){1}{\oval(4.0,4.0)[r]}
\multiput(50,24)(0,8){1}{\oval(4.0,4.0)[l]}
\put(50,36){\circle{12.}}
\put(30,10){\line(2,1){15.}}
\put(30,9){\line(2,1){15.}}
\put(55,17.5){\line(2,-1){15.}}
\put(55,16.5){\line(2,-1){15.}}
\put(50,-10){\makebox(0,0){\small e)}}
\end{picture}
\end{tabular}
\vspace{0.5cm}
\caption{
\protect\it
Feynman diagrams contributing to the Born and the radiative correction
cross sections.
}
\label{feyn}
\end{figure}

In this paper we consider the case of elastic
measurement
\begin{equation}\label{process}
e(k_1)+p(p_1) \rightarrow e(k_2)+p(p_2)
\end{equation}
where the initial electron is polarized longitudinally and only momentum
of the
final proton is measured or is used in $Q^2=-(p_2-p_1)^2$ reconstruction.
The corresponding method is known as reconstruction
within hadronic variables \cite{Bardin,Blu}.
 There are two types of
such experiments. 
 In the first type
the proton is
considered
to be polarized either longitudinally or transversely. In the second case
the
polarization states of recoil proton are measured. We calculate radiative
corrections (see Fig.\ref{feyn}) for these measurements and for hadronic
variables
reconstruction method.

Measurement of final proton momentum allows to define three
independent kinematic
variables. A natural choice for them are $Q^2$, azimuthal angle and
so-called inelasticity $u=(k_1+p_1-p_2)^2-m^2$, where $m$ is the electron 
mass.
 Since at
the Born
level $u=0$ and $u>0$ if the additional photon is emitted in the final
state of (\ref{process}), one can constrain (or cut) the range of $u$ 
to reduce value
of radiative corrections (RC).

We calculate RC to the cross sections and asymmetries differential
in $Q^2$. When leptonic variables reconstruction method is used, the
integration over photonic momentum phase space cannot be performed
analytically. The reason is that an argument of formfactors ($Q^2$
transferred
to
the proton) depends on the photon momentum. So the integration is left
for
numerical analysis in order to avoid additional assumptions about
specific models for formfactors. The advantage of the 
reconstruction method 
in terms of
hadronic variables
is that the integration can be performed
analytically and completely both for unpolarized and polarized parts of
the
cross section. A more detailed discussion of advantages of RC calculation
for
elastic
scattering
within leptonic and hadronic variables can be found in ref.\cite{AAM}

There are two possibilities in considering the polarization parts of the
cross section. We consider different polarization states. It corresponds
to different coefficients of expansion of polarization vectors, while the
form of the expansion is the same. Since these coefficients depend on
inelasticity (see Section 2), which is a function of photon momentum, we
can produce either
formulae for all specific polarization cases or get only one set of
formulae for the polarization states but leaving integration over
inelasticity for numerical analysis. We choose the second way. We will see
in Section 3, that obtained formulae are very simple even in the
general case.

We use the approach of Bardin and Shumeiko \cite{BSh} for this
calculation.
Advantages of the approach are absence of any approximations both in
the procedure of extraction and cancellation of infrared divergent terms
and
in integration over photon phase space. However it is known that mass
singularity terms still remain within these formulae. In our case we
perform additional procedure to extract these mass singularity terms.
The result of the expansion can be presented as
\begin{equation}\label{expansion}
\sigma=\log\frac{Q^2}{m^2} A(Q^2) + B(Q^2) + O\biggl(\frac{m^2}{Q^2}\biggr)
\end{equation}
where independent on electron mass functions $A(Q^2)$ and $B(Q^2)$ appear as LO
and NLO contribution to the cross section. The explicit formulae in the form of
(\ref{expansion}) are presented in Section 3.

Section 4 is devoted to numerical analysis within kinematic conditions of
experiments at JLab. We analyze RC to unpolarized cross section and
different
asymmetries accessed in polarization measurements. Furthermore, obtained results
can be applied for other reactions, where 
$ep\rightarrow ep\gamma$
is a 
background process. In Section 4 a specific example is considered when the
process
is background to measurements of pion production by
polarized real photons at JLab.

\section{The Born cross section and kinematics}

The Born cross section of the process (\ref{process})
can be written in the form
\begin{equation}\label{sigma0}
d\sigma_0=\frac{M_0^2}{4pk_1}d\Gamma_0=M_0^2\frac{dQ^2}{16\pi S^2}
\end{equation}
where $S=2k_1p$. Kinematic limits for $Q^2$ are defined as
\begin{equation}\label{q2minmax}
0\leq Q^2 \leq \frac{\lambda_s}{S+m^2+M^2}, \qquad \lambda_s=S^2-4m^2M^2
\end{equation}
where ($m,M$ are the electron and proton masses). Here, for 
generality, we keep the electron mass, however, below we will neglect it
wherever
 possible. At the Born level the mass can be set to zero in the final
formulae, while for RC cross section we will construct expansion
(\ref{expansion}). We note that in general  we should keep explicit
azimuthal angular dependence. The cross section (both Born and RC ones)
does not
depend on this angle, but kinematic cuts can. We assumed that there is 
only a
cut on inelasticity or all other cuts can be effectively reduced to it. In this
case integration over the azimuthal angle can be performed analytically, that
results to (\ref{sigma0}).
Born matrix element is the contraction of leptonic and hadronic tensors:
\begin{equation}\label{convol}
M^2_0=\frac{e^4}{Q^4}L^0_{\mu\nu}W_{\mu\nu}
\end{equation}
We use standard definitions for the leptonic tensor,
\begin{equation}
L_{\mu\nu}^0=\frac{1}{2}{\rm Tr} (\hat k_2+m)\gamma_\mu
(1+\gamma_5\hat\xi)(\hat k_1+m)\gamma_\nu.
\end{equation}
The hadronic tensor can be described by four structure functions in all
considered
cases:
\begin{equation}
W_{\mu\nu}=\sum_i w_{\mu\nu}^i {\cal F}_i
\end{equation}
with
\begin{eqnarray}
w_{\mu\nu}^1=-g_{\mu\nu} \qquad
w_{\mu\nu}^2=\frac{p_{\mu}p_{\nu}}{M^2} \qquad
w_{\mu\nu}^3=-i\epsilon_{\mu\nu\lambda\sigma}{q_{\lambda}\eta_{\sigma}
\over M}, \qquad
w_{\mu\nu}^4=i\epsilon_{\mu\nu\lambda\sigma}{q_{\lambda}p_{\sigma}\;
\eta q \over M^3}
\end{eqnarray}
and ($\tau_p=Q^2/4M^2$)
\begin{eqnarray}
{\cal F}_1=4 \tau_p M^2 G_m^2 \qquad
{\cal F}_2=4 M^2 \frac{G_E^2+\tau_p G_m^2}{1+\tau_p}\qquad
{\cal{F}}_3=-2M^2G_EG_M, \qquad
{\cal{F}}_4=-M^2G_M{G_E-G_M\over 1+\tau_p}.
\end{eqnarray}

Polarization effects are described by polarization four-vectors of the lepton
($\xi$)
and proton ($\eta$).
Following \cite{ASh} we expand them
over the measured momenta $k_1$, $p_1$ and $p_2$
\begin{eqnarray}
 \xi&=&{2\over \sqrt{\lambda_s} }\bigl({S \over 2m} k_1-mp_1\bigr)
 \nonumber\\
 \eta&=&2a_{\eta}k_1+b_{\eta}q+c_{\eta}(p_1+p_2)
\end{eqnarray}
where $q=p_1-p_2$. In this paper we consider four\footnote{The considered 
model independent RC does not include box-type contributions and therefore 
cannot lead to the additional T-odd polarization and/or asymmetry.
The polarization part of the cross section of the model independent RC
as well as of
the
Born cross section is exactly zero for normal polarization} polarization
states, when
initial or final protons are polarized along or perpendicular (in the scattering
plane) to the vector $\vec q$. For these cases the following expressions of the
coefficients are used,
\begin{equation}\label{etal}
 a_{\eta}^{L}=0, \qquad
 b_{\eta}^{L}={Q^2+4M^2\over 2M\sqrt{\lambda_M}}, \qquad
 c_{\eta}^{L}={Q^2\over 2M\sqrt{\lambda_M} }
\end{equation}
and
\begin{eqnarray}\label{etat}
 a_{\eta}^{T}   ={Q^2(Q^2+4M^2)\over 2\sqrt\lambda_h \sqrt{\lambda_M}}, \qquad
 b_{\eta}^{T}    ={(Q^2+4M^2)Q^2_u\over 2\sqrt\lambda_h  \sqrt{\lambda_M}},   \qquad
 c_{\eta}^{T}    =-{Q^2(2S-Q^2_u)\over 2\sqrt\lambda_h  \sqrt{\lambda_M}}.
\end{eqnarray}
Here $Q^2_u=Q^2+u$, $\lambda_h=SQ^2(S-Q^2_u)-M^2Q^4_u-m^2\lambda_M$
and $\lambda_M=Q^2(Q^2+4M^2)$. These formulae are given for the case of initial
particles polarizations. For the final polarizations these formulae can
be kept but $b^L_{\eta}\rightarrow -b^L_{\eta}$. One can make sure that
initial and
final polarization vectors  exactly satisfy the necessary conditions:
\begin{equation}
\eta_{L,T}^2=-1, \qquad \eta_L\eta_T=0, \qquad q\eta_T=0, \qquad
{\bar p} \eta_{L,T}=0
\end{equation}
where $\bar p$ is $p_1$ or $p_2$ in dependence of initial or final
polarization vector
is considered.

Calculating the contractions in (\ref{convol}) we obtain the Born cross
section
in the form
\begin{equation}
\frac{d\sigma_0}{dQ^2}={2\pi\alpha^2 \over S^2 Q^4} \sum_i
\theta_B^i
{\cal F}_i,
\label{si0}
\end{equation}
where
\begin{eqnarray}
\theta_1^0 &=& 2Q^2
\nonumber \\
\theta_2^0 &=& {1\over M^2} (S^2-Q^2S-M^2Q^2)
\nonumber \\
\theta_3^0 &=& -{2Q^2\over M} (Q^2 a_{\eta}+(2S-Q^2)c_{\eta})
\nonumber \\
\theta_4^0 &=& {Q^4\over M^3} (2S-Q^2) (a_{\eta}-2b_{\eta})
\nonumber
\end{eqnarray}

\section{Radiative corrections}

 The cross section of radiative process (Fig. \ref{feyn}b,c)
\begin{equation}
 e(k_1) + p (p_1) \longrightarrow e'(k_2) + \gamma(k) + p(p_2),
\end{equation}
can be presented in a general form
\begin{equation}
d\sigma_r=\frac{M_r^2}{4pk_1}d\Gamma_r
\end{equation}
The
 phase space
of three final particles
\begin{equation}
d\Gamma_r=\frac{1}{(2\pi)^5}
\frac{d^3p_2}{2p_{20}}
\frac{d^3k_2}{2k_{20}}
\frac{d^3k}{2k_{0}} \delta(p+k_1-k_2-k-p_2)
\end{equation}
can be parameterized in terms of four invariant variables
\cite{BarKal,AISh2}: $Q^2$,
$u=2k_2k$, $w=2k_1k$ and $z_2=2p_2k$
\begin{equation}\label{eq70}
d\Gamma_r={dQ^2\over 8(2\pi)^4S}
\int\limits_0^{u_m} {du}
\int\limits_{w_{min}}^{w_{max}} {dw}
\int\limits_{z_{min}}^{z_{max}} \frac{dz_2}{\sqrt{R_z}}
\end{equation}
where $R_z$ comes from Gram determinant and coincides with a
standard $R_z$-function from Bardin-Shumeiko approach
\cite{BSh,BarKal}. The formalism to calculate it is general and is
developed in
\cite{Bukling}.
Explicit formulae for analytical integration are presented in papers
\cite{BarKal,AISh2,AAM}. Note
that
integration
over the azimuthal angle is assumed to be performed (see discussion after
Eq.(\ref{q2minmax})).

The matrix element squared of radiated process reads
\begin{equation}
M^2_r=\frac{e^6}{Q^4_h}L^r_{\mu\nu}W_{\mu\nu}
\label{r1}
\end{equation}
The leptonic tensors of radiative process
in (\ref{r1}) looks like
\begin{equation}
 L^r_{\mu \nu}=-\frac 12 Tr \; 
(\hat {k}_2+m)
\Gamma _{\mu \alpha}
(1+\gamma_5 \hat{\xi})
(\hat {k}_1+m)
\bar{\Gamma }_{\alpha \nu}
,
\label {cl1}
\end {equation}
where
\begin {eqnarray}
 \Gamma _{\mu \alpha}=
\left [\left(\frac {k_{1 \alpha}}{kk_1}- \frac {k_{2 \alpha}}{kk_2}
\right)\gamma_{\mu}-
\frac {\gamma_{\mu}\hat{k}\gamma_{\alpha}}{2kk_1}-
\frac{\gamma_{\alpha}\hat{k}\gamma_{\mu}}{2kk_2} \right],
 \bar{\Gamma}_{\alpha \nu}=
\left [\left(\frac {k_{1 \alpha}}{kk_1}- \frac {k_{2 \alpha}}{kk_2}
\right)\gamma_{\nu}
-\frac{\gamma_{\alpha}\hat{k}\gamma_{\nu}}{2kk_1}
-\frac {\gamma_{\nu}\hat{k}\gamma_{\alpha}}{2kk_2}
 \right].
\label{g12}
\end {eqnarray}

The calculation of the cross section of the radiative process is not so
straightforward because of infrared divergence.
The separation of the infrared divergence can be performed in a
standard way \cite{BSh}:
\begin{equation}\label{diff}
d\sigma _r=d\sigma _r^{IR}+d\sigma _r
-d\sigma _r^{IR}=
d\sigma _r^{IR}+d\sigma _r^F
.
\end{equation}
The first term in r.h.s. of Eq. (\ref{diff}) has an infrared divergence
and can be presented in the form:
\begin{equation}\label{ikr}
d\sigma _r^{IR}
=\frac {\alpha}{\pi}\delta ^{IR}d\sigma _0
=\frac {\alpha}{\pi}(\delta_S+\delta_H)d\sigma _0
\end{equation}
while the second one in (\ref{diff}) is finite for $k\rightarrow
0$.
The quantities $\delta_{S}$ and $\delta_{H}$ appear after additional
splitting the integration region over inelasticity $u$ by the
infinitesimal parameter $\bar u$
\begin{eqnarray}
\delta_{S}&=&\frac{-1}{\pi}\int\limits_0^{\bar
u}du\int\frac{d^{n-1}k}{(2\pi\mu)^{n-4}k_0}F_{IR}\delta((\Lambda_h-k)^2-m^2)
\nonumber\\
\delta_{H}&=&\frac{-1}{\pi}\int\limits_{\bar
u}^{u_m}du\int\frac{d^3k}{k_0}F_{IR}\delta((\Lambda_h-k)^2-m^2)
\end{eqnarray}
where $\Lambda_h=k_1+p_1-p_2$ and
\begin{equation}
F_{IR}=
\frac {Q^2}{uw}-\frac {m^2}{u^2}-\frac
{m^2}{w^2}.
\label{fir}
\end{equation}
For this specific procedure it is convenient to keep integration over
photon
momentum as it is. In this case one delta function appears in
the integrand. It is equivalent to integration over $w$ and $z_2$ in
(\ref{eq70}).
The integration gives the following results
\begin{eqnarray}\label{deltash}
\delta_S&=&2\biggl(P^{IR}+\log\frac{\bar
u}{m\mu}\biggr)(l_m-1)+1+l_m-l_m^2
-\frac{\pi^2}{6},
\nonumber\\
\delta_H&=&2(l_m-1)\log\frac{u_m}{\bar u}-{1\over 2}
(l_v+l_m)^2+l_v+l_m
-l_w(l_m+l_w-l_v)
-\frac{\pi^2}{6}-{\rm Li}_2\biggl(-{u_m
\over Q^2}\biggr),
\nonumber\\&&
l_m=\log\frac{Q^2}{m^2}, \qquad
l_v=\log\frac{u_m}{Q^2}, \qquad
l_w=-\log(x_m),\qquad x_m=\frac{Q^2}{Q^2+u_m}.
\end{eqnarray}

The explicit expression for
the additional photon exchange contribution coming from diagram (\ref{feyn}d)
has also the factorized form:
\begin{equation}\label{vvkl}
d\sigma _V=\frac {\alpha}{\pi}\delta ^{V}d\sigma _0
\end{equation}
where
\begin{equation}
\delta ^V=-2\left ( P^{IR}+\log \frac {m}{\mu} \right)(l_m-1)
-\frac 12 l_m^2+\frac 32 l_m -2+\frac{\pi ^2}6
\end{equation}
Thus the sum of contributions (\ref{ikr}) and (\ref{vvkl})
\begin{equation}
\delta ^{el}=
\delta ^{V}+
\delta _S+
\delta^{IR}_H=l_m(l_v-l_w+\frac{3}{2})-l_v
-1-\frac{3}{2}l_w^2-\frac{1}{2}l_v^2+2l_wl_v
-{\rm Li}_2(x_m)
\end{equation}
is infrared free.

The second term in (\ref{diff}) is infrared free.
 Hence this contribution can be written using the expression for
photon phase space (\ref{eq70}) as
\begin{equation}
d\sigma ^F =\frac{\alpha ^3}{4} \frac {d Q^2}{S^2Q^4}
\sum_{i=1}^{4}\theta _i^F{\cal F}_i.
\end{equation}
where only coefficients $\theta _i^F$ include integrals:
\begin{equation}\label{integr}
\theta _i^F=
\frac 1{\pi }
\int\limits_0^{u_m} {du}
\int\limits_{w_{min}}^{w_{max}} {dw}
\int\limits_{z_{min}}^{z_{max}} \frac{dz_2}{\sqrt{R_z}}
[L^r_{\mu \nu}-\alpha F_{IR}L^0_{\mu \nu}]w_{\mu \nu}^i
=\frac 1{\pi }
\int\limits_{x_m}^1 \frac {Q^2}{x^2}{dx}
\int\limits_{w_{min}}^{w_{max}} {dw}
\int\limits_{z_{min}}^{z_{max}} \frac{dz_2}{\sqrt{R_z}}
[L^r_{\mu \nu}-\alpha F_{IR}L^0_{\mu \nu}]w_{\mu \nu}^i.
\end{equation}
Here we use the transformation of external variable $u$,
\begin{equation}
x={Q^2\over u+Q^2},
\end{equation}
to have final results in
the simplest form, where  $x_m$ is defined in Eq.(\ref{deltash}).

Integrals in (\ref{integr}) can be calculated analytically completely.
However, as
we discussed in Introduction, one integration in  polarization part of the
cross
section is retained for generality:

\begin{eqnarray}
\theta _1^F&=&2Q^2[\frac
1{x_m}(2L-1)+L+2l_m(l_w-3)-2l_w+l_w^2
-2{\rm Li}_2(x_m)+\frac {\pi ^2}3
]
\\
\theta _2^{F}&=&\frac 1{M^2}
\Bigl\{
Q^4l_w-Q^2(S+M^2)[
L+2l_m(l_w-3)+l_w^2
-2{\rm Li}_2(x_m)+\frac {\pi ^2}3
]
+\frac{M^2Q^2}{x_m}[1-2L]
\nonumber\\&&
+S(Q^2-S)[2x_m(1-L)-4l_w-1]
+S^2[x_m^2L+3-x_m-2x_m^2-6l_m]
\Bigr\}
\nonumber\\
\theta _3^{F}&=&\frac {2Q^2}M
\int \limits_{x_m}^1\frac {dx}x\Bigl[
Q^2a_{\eta }h_1+(2Sh_2+Q^2h_3)c_{\eta }+(Q^2\Delta a_{\eta }
+(2S-Q^2)\Delta c_{\eta })h_4
\Bigr]+\theta_3^0\delta_{pol}^F
\nonumber\\
\theta _4^{F}&=&\frac{Q^4}{M^3}
\int \limits_{x_m}^1\frac {dx}x\Bigl[
(Q^2h_1+2Sh_3)a_{\eta }+
(2Sh_2+Q^2h_3)b_{\eta }+
(2S-Q^2)h_4(\Delta b_{\eta }-\Delta a_{\eta })
\Bigr]+\theta_4^0\delta_{pol}^F
\nonumber\end{eqnarray}
Here the contribution of polarization part is presented in such a form where
$x$ (or $u$) dependence is included only in polarization coefficients and
in 
functions:
\begin{eqnarray}
h_1&=&(3x^2-x+5-2l_x(2x^2+x+1))/x^2,
\\
h_2&=&1-8x-2l_x(1-x),
\nonumber\\
h_3&=&(5x-5+2l_x(x+1))/x,
\nonumber\\
h_4&=&(3x+4-4l_xx)/(1-x).
\nonumber\end{eqnarray}
The special procedure of additional subtraction was applied for polarization
contribution in order to extract mass singularity terms to $l_m$, which is
contained in $l_x=l_m-\log(x(1-x))$ and $L=l_m-l_v+2l_w$ now. In
subtracted (and
added)  terms arguments of polarization coefficients were taken for $x=1$ (or
$u=0$) as for Born process. As a result differences $\Delta \{a,b,c\}_{\eta
}=\{a,b,c\}_{\eta }(x)-\{a,b,c\}_{\eta}(1)$ appeared in the integrands. Because
of added terms does not contain $x$-dependence of polarization
coefficients, they
can be integrated explicitly. This is the origin of factorized terms with
$\delta_{pol}^F$ in expressions for $\theta _3^{F}$ and $\theta _4^{F}$:
\begin{eqnarray}
\delta _{pol}^F&=&-\int \limits^1_{x_m}dx\Bigl[
\frac{x^4 Q^2(3Q^2(1-x)+4xm^2)}{(Q^2(1-x)+m^2x)^2}
+3x^3+3x^2+3x+3 \Bigr]
\\
&=&-3(l_m+l_v-l_w)-1
\end{eqnarray}

The last  contribution which has to be taken into account (see Fig. \ref{feyn}e)
is vacuum
polarization by leptons and by hadrons ($d\sigma_{vac}$).
Leptonic contribution comes from QED
(see \cite{ASh}, for example), while hadronic contribution can be obtained from
experimental data for the process $e^+e^- \rightarrow hadrons$. We will use
a model developed in ref.\cite{vachadr}. The vacuum polarization
correction
has also a
factorized form and we denote them $\delta_{vac}^l$ and $\delta_{vac}^h$ for
leptonic and hadronic contributions.

The final result for the RC cross section is obtained by adding all the
considered contributions:
\begin{equation}\label{res}
d\sigma _V+
d\sigma _r+d\sigma_{vac}
=\frac{\alpha ^3}{4} \frac {d Q^2}{S^2Q^4}
\sum_{i=1}^{4}[\theta _i^F+4(\delta ^{el}+\delta_{vac}^l+\delta_{vac}^h)
\theta _i^0]{\cal F}_i.
\end{equation}

This expression is the correction to the cross section differential in
$Q^2$ only.
Sometimes it is necessary to have formulae for the cross section versus
inelasticity
also. They can be obtained by straightforward differentiation over $x_m$
(or
$u_m$) of the result (\ref{res}):
\begin{equation}
\frac{d\sigma _r}{dx}
=\frac{d\sigma _r}{du} {Q^2\over x^2}
=\frac{\alpha ^3}{4} \frac {d Q^2}{S^2Q^4}
\sum_{i=1}^{4}
\frac {d\theta _i^R}{dx}
{\cal F}_i.
\end{equation}
where
\begin{eqnarray}\label{eq43}
\frac{d\theta ^{R}_1}{dx}&=&-
2Q^2\frac {2(1+x^2)L+1-8x}{(1-x)x^2}
\\
\frac{d\theta ^{R}_2}{dx}&=&\frac 1{M^2}
\Bigl\{
\frac{4(x^2S(xS-Q^2)-M^2Q^2)}{(1-x)x}
-\frac {Q^4}{x}
\nonumber\\
&&
-\frac{(xS(xS-Q^2)-M^2Q^2)[2(1+x^2)L+1-2x(1+x)]}{(1-x)x^2}
\Bigr\}
\nonumber\\
\frac{d\theta ^{R}_3}{dx}&=&
\frac {2Q^2(2(1+x^2)L-8x^2+6x-5)
(Q^2a_{\eta }+(2xS-Q^2)xc_{\eta })}{(1-x)x^3M}
\nonumber\\
\frac{d\theta ^{R}_4}{dx}&=&
\frac {Q^4(2(1+x^2)L-8x^2+6x-5)
(2xS-Q^2)(xb_{\eta }-a_{\eta })}{(1-x)x^3M^3}
\nonumber\end{eqnarray}
It is clear that only hard photon radiation can contribute to the
functions. The terms containing $x_m$ in $\delta 's$ and in subtracted
parts
$d\sigma_r^{IR}$ have to be completely canceled. It can be seen in
polarization parts that the terms containing  $a_{\eta}(1)$, $b_{\eta}(1)$
and $c_{\eta}(1)$ explicitly cancel out
 in final expressions (\ref{eq43}).

\section{Numerical analysis}

\begin{figure}[!t]
\unitlength 1mm
\begin{center}
\begin{picture}(160,80)
\put(0,-10){
\epsfxsize=8cm
\epsfysize=8cm
\epsfbox{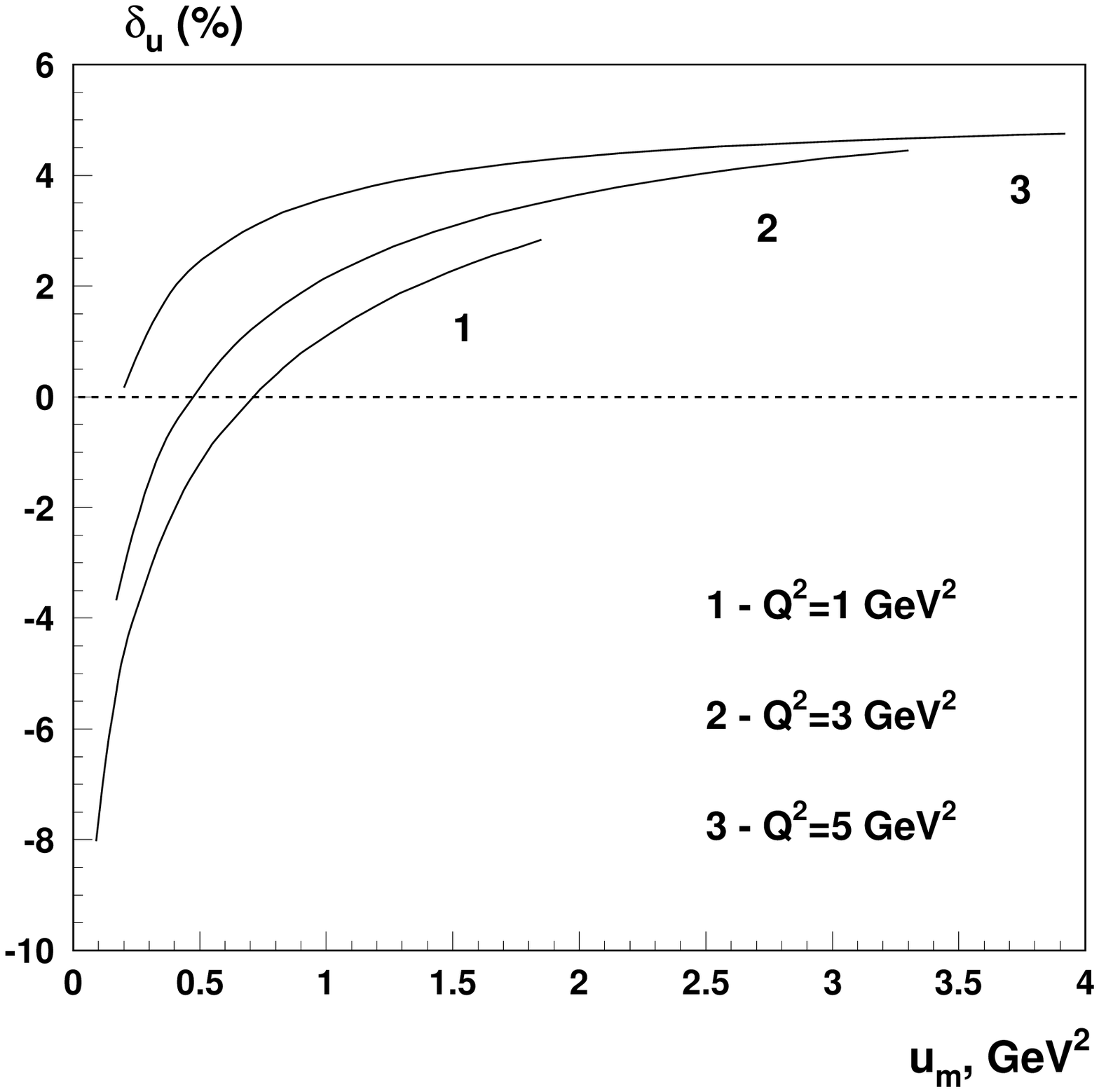}
}
\put(80,-10){
\epsfxsize=8cm
\epsfysize=8cm
\epsfbox{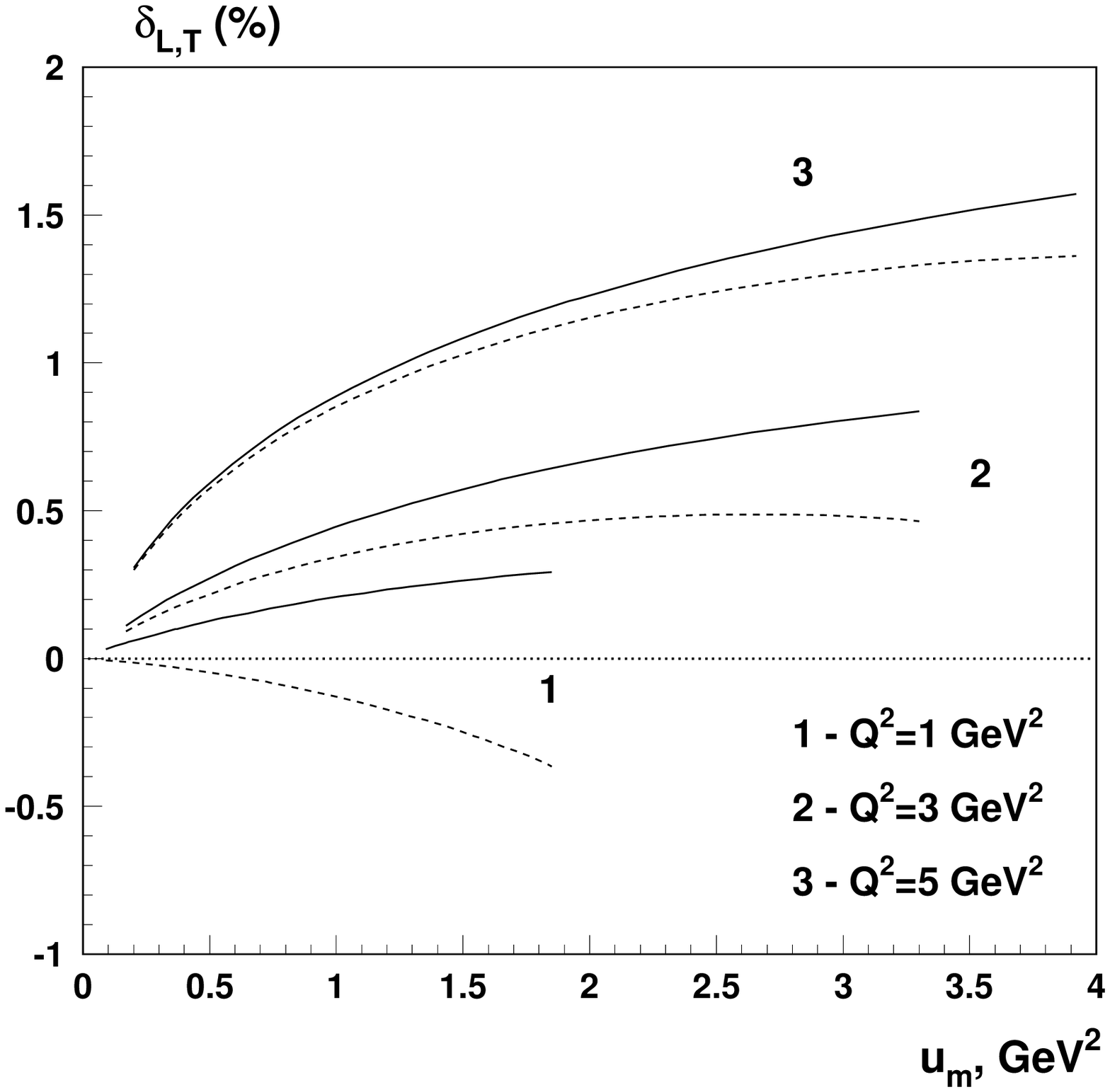}
}
\end{picture}
\end{center}
\caption{\label{fig1}
Radiative corrections to the unpolarized cross section (left plot) and
polarization asymmetries (right plot) defined in (\protect\ref{deltas}).
Solid and dashed  lines corresponds to
longitudinal and transverse cases. $S$=8 GeV$^2$.
}
\end{figure}

The obtained results can have direct application in experiments at JLab, so we
perform the numerical analysis within kinematic conditions of JLab. First 
we consider the unpolarized cross section and RC to it. The next step is
consideration of polarization effects in two cases when initial or final protons
are polarized.

Both the Born and observed cross sections can be split into unpolarized
and
polarized parts:
\begin{equation}
{d\sigma^{born,obs} \over dQ^2}=\sigma_u^{b,obs} \pm \sigma_p^{b,obs}
\end{equation}
We consider four different polarization states described in
Eqs.(\ref{etal},\ref{etat}), so
we have four different polarized parts of cross sections, which, of course,
corresponds to only one unpolarized cross section. Let us define the relative
corrections to the observable quantities in the current experiments:

\begin{eqnarray}\label{deltas}
&&
\delta_u={\sigma_u^{obs} \over \sigma^b_u} -1, \qquad
\delta_{L,T}=\left[ \frac{\sigma_p^{obs}}{\sigma_u^{obs}}
-\frac{\sigma_p^{b}}{\sigma_u^{b}} \right]
\left[ \frac{\sigma_p^{b}} {\sigma_u^{b}} \right]^{-1}
={\sigma_u^{b} \over \sigma^{obs}_u} {\sigma_p^{obs} \over \sigma^{b}_p}-1
\nonumber  \\[0.4cm] &&
\delta_r=\left[ \frac{\sigma_{pT}^{obs}}{\sigma_{pL}^{obs}}
-\frac{\sigma_{pT}^{b}}{\sigma_{pL}^{b}} \right]
\left[ \frac{\sigma_{pT}^{b}} {\sigma_{pL}^{b}} \right]^{-1}
={\sigma_{pL}^{b} \over \sigma^{obs}_{pL}}
{\sigma_{pT}^{obs} \over \sigma^{b}_{pT}}-1
\end{eqnarray}

The first correction $\delta_u$ is the relative correction to unpolarized
cross section.
The $\delta_{L,T}$ are corrections to polarization asymmetries measured by rotating
the polarization states of initial protons. At last, the quantity
$\delta_r$ is the
correction to the measured ratio of final proton polarizations
\cite{mjones,jlab2}. The
correction to the unpolarized cross section is presented in Figure
\ref{fig1}a. The
behavior is quite typical. For the very hard inelasticity cut ($u_m \ll
Q^2$) the
positive contribution due to real bremsstrahlung is suppressed, 
so there is only
negative loop correction contributing to cross section. 
Different ending values for
the curves corresponds to different kinematically allowed regions.

The correction to the asymmetry is given in the plot
\ref{fig1}b. The
magnitude of the correction does not exceed 1.5\%. It goes up with
increasing
of $Q^2$ and
inelasticity cut. Second effect is clear because the only unfactorized (different
for unpolarized and polarized parts) hard bremsstrahlung can contribute to RC to
asymmetries.  The $Q^2$-dependence is also understood because of contribution of
a large logarithm $\log(Q^2/m^2)$. Both the cross sections    and
asymmetries have
non-zero leading contributions, so the correction is larger when $Q^2$ is going
up.

In the case of calculation of correction to the transferred polarization
experiment the
correction is negative and does not exceed 1\%. It is in agreement with our
earlier estimation of the effect \cite{AAM1,AAM}.

\begin{figure}[!t]
\unitlength 1mm
\begin{center}
\begin{picture}(160,80)
\put(60,-10){
\epsfxsize=8cm
\epsfysize=8cm
\epsfbox{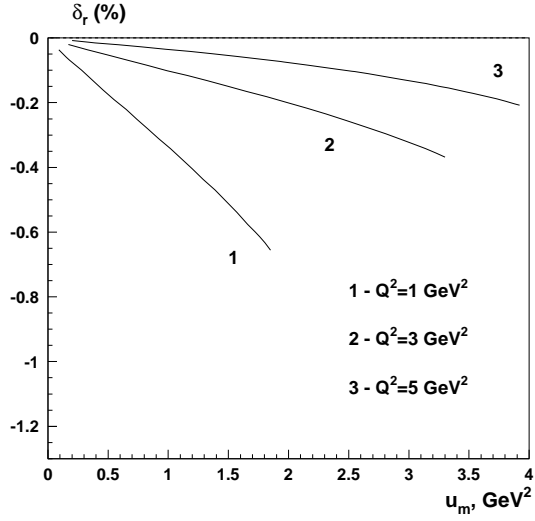}
}
\end{picture}
\end{center}
\caption{\label{fig2}
Radiative correction to recoil proton polarization for (\protect\ref{proe}) in
the region the invariant mass of the unobserved state   close to the pion
mass.
$S$=8GeV$^2$.
}
\end{figure}

\begin{figure}[!t]
\unitlength 1mm
\begin{center}
\begin{picture}(160,80)
\put(0,-10){
\epsfxsize=8cm
\epsfysize=8cm
\epsfbox{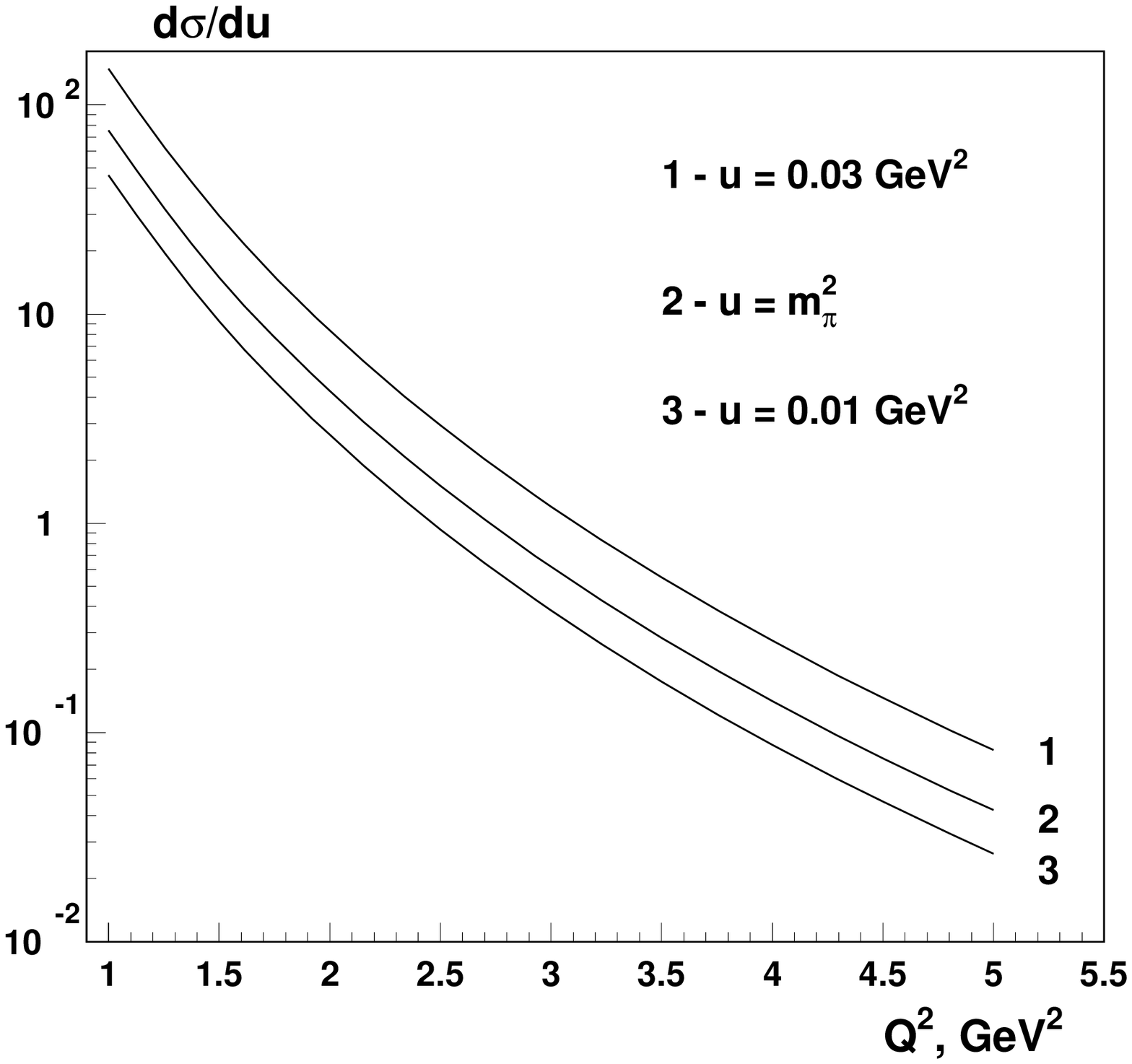}
}
\put(80,-10){
\epsfxsize=8cm
\epsfysize=8cm
\epsfbox{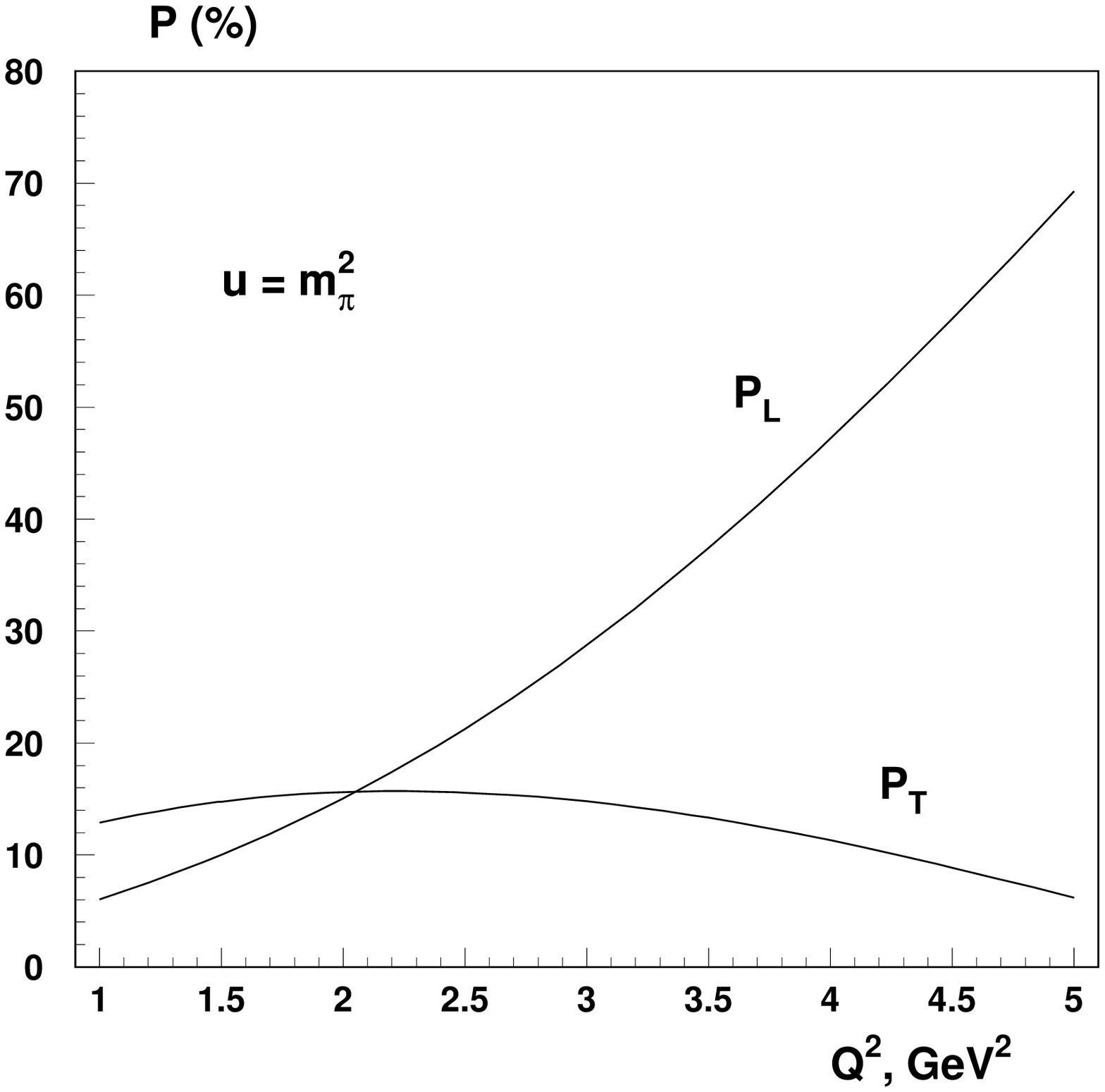}
}
\end{picture}
\end{center}
\caption{\label{fig3}
The cross section and proton polarizations of the process
(\protect\ref{proe}). $Q^2$ here is equivalent to $-t$ defined in
\cite{piexp}.
}
\end{figure}

An interesting application of our result can be found for the
process
\begin{equation}\label{propi}
\vec \gamma + p \rightarrow \pi^0 + \vec p
\end{equation}
recently analyzed in JLab experiment \cite{piexp}.
Usually the photons in the reaction are produced by polarized electron beam. As
a result both photon and electrons are in the beam. 
Therefore when invariant mass of undetected $e$ and $\gamma$ is close to
the pion mass $m_{\pi}$ the process considered in this paper
\begin{equation}
\label{proe} \vec e + p \rightarrow e + \gamma + \vec p
\end{equation}
can be background  to the process (\ref{propi}).
In the end of the section we analyze unpolarized cross
section and recoil proton polarization due to the process (\ref{proe}).

In Figure \ref{fig3}a we show the cross section vs $Q^2$ for several values of
the unobserved invariant masses set close to the pion mass. Plot
\ref{fig3}b shows polarization of protons due to the bremsstrahlung
process (\ref{proe}) being 
background to the measurements \cite{piexp}.

\section{Conclusion}

In this paper we consider the radiative effects in elastic electron-proton
scattering with the hadronic variables reconstruction method. Within this
method the information on proton final momentum is used to reconstruct the 
kinematic variable $Q^2$. Another kinematic variable which can be
also reconstructed from the measured recoil momentum of proton is
inelasticity. The Born cross
section does not depend on the inelasticity, so this fact may be used in 
measurements to make
a cut on this variables.
 It allows to reduce radiative corrections
essentially, but at the cost of significant loss in
statistics. In this paper we obtain explicit formulae for the cross
section and spin asymmetries versus $Q^2$ and inelasticity $u$. 
We calculated RC to unpolarized cross section and polarization observables in
kinematic conditions of
experiments held at JLab.
We found that correction is about one per cent. 
It increases when both $Q^2$
and inelasticity go up. Also we calculated a cross section and 
proton polarization from elastic ep-events that produce background  
 in neutral pion production
by polarized real photon at JLab.

\section*{Acknowledgements}
We thank our colleagues at Jefferson Lab for useful discussions.
AA and IA thank the US Department of Energy for support under contract
DE-AC05-84ER40150. Work of NM was in addition supported by Rutgers
University
through NSF grant PHY 9803860 and Ukrainian DFFD.

\end{document}